\documentclass[twocolumn,showpacs,aps,prl,superscriptaddress,floatfix,letter]{revtex4}

\usepackage{graphicx}
\usepackage{epsfig}
\usepackage{amsmath}
\usepackage{amssymb}
\usepackage{hhline}
\usepackage{multirow}
\usepackage{fancyheadings}
\usepackage{ifthen}

\RequirePackage{xspace}

\newcommand{\BABARPubYear}    {05}
\newcommand{\BABARPubNumber}  {049}

\newcommand{\SLACPubNumber}   {11582}
\newcommand{\LANLNumber}      {0601046}

\usepackage{relsize}
\newcommand {\collab}{Collaboration,}
\def\babar{\mbox{\slshape B\kern-0.1em{\smaller A}\kern-0.1em
    B\kern-0.1em{\smaller A\kern-0.2em R}}}
\def\geantfour      {\mbox{\tt GEANT4}\xspace}
\def\jetset     {\mbox{\tt JETSET \hspace{-0.5em}7\hspace{-0.3em}.\hspace{-0.2em}4}}
\def\pep2{PEP-II}
\def\to                 {\ensuremath{\rightarrow}\xspace}
\def\B       {\ensuremath{B}\xspace}
\def\Bbar    {\kern 0.18em\overline{\kern -0.18em B}{}\xspace}
\def\Bb      {\ensuremath{\Bbar}\xspace}
\def\BB      {\ensuremath{B\Bbar}\xspace} 
\def\Bz      {\ensuremath{B^0}\xspace}
\def\Bzb     {\ensuremath{\Bbar^0}\xspace}
\def\Vts  {\ensuremath{|V_{ts}|}\xspace}
\def\Vub  {\ensuremath{|V_{ub}|}\xspace}
\def\Vcb  {\ensuremath{|V_{cb}|}\xspace}
\def\Y#1S{\ensuremath{\Upsilon{(#1S)}}\xspace}
\def\FourS {\Y4S}
\def\upsbb   {\ensuremath{\FourS \to \BB}\xspace}
\def\CP                {\ensuremath{C\!P}\xspace}
\def\nub        {\ensuremath{\overline{\nu}}\xspace}
\def\KL    {\ensuremath{K^0_{\scriptscriptstyle L}}\xspace} 
\def\KS    {\ensuremath{K^0_{\scriptscriptstyle S}}\xspace} 
\def\Dstar   {\ensuremath{D^*}\xspace}

\newcommand{\tev}{\ensuremath{\mathrm{\,Te\kern -0.1em V}}\xspace}
\newcommand{\gev}{\ensuremath{\mathrm{\,Ge\kern -0.1em V}}\xspace}
\newcommand{\mev}{\ensuremath{\mathrm{\,Me\kern -0.1em V}}\xspace}
\newcommand{\kev}{\ensuremath{\mathrm{\,ke\kern -0.1em V}}\xspace}
\newcommand{\ev}{\ensuremath{\mathrm{\,e\kern -0.1em V}}\xspace}
\newcommand{\gevc}{\ensuremath{{\mathrm{\,Ge\kern -0.1em V\!/}c}}\xspace}
\newcommand{\mevc}{\ensuremath{{\mathrm{\,Me\kern -0.1em V\!/}c}}\xspace}
\newcommand{\gevcc}{\ensuremath{{\mathrm{\,Ge\kern -0.1em V\!/}c^2}}\xspace}
\newcommand{\mevcc}{\ensuremath{{\mathrm{\,Me\kern -0.1em V\!/}c^2}}\xspace}
\def\ps   {\ensuremath{\rm \,ps}\xspace}

\newcommand {\Bxulnu}{\ensuremath{\Bb \rightarrow X_u \ell \bar{\nu}}}
\newcommand {\Bxclnu}{\ensuremath{\Bb \rightarrow X_c \ell \bar{\nu}}}
\newcommand {\Bxlnu}{\ensuremath{\Bb \rightarrow X \ell \bar{\nu}}}
\newcommand {\Btaunu}{\ensuremath{\Bb\!\rightarrow\!X \tau \bar{\nu}_{\tau}}}
\newcommand {\Bsg}{\ensuremath{B \!\rightarrow\! X_s \gamma}}
\newcommand {\mX}{\ensuremath{m_{X}}}
\newcommand{\LLRc}{\ensuremath{\zeta}}
\newcommand {\mXmax}{\LLRc}
\newcommand{\dGc}{\ensuremath{\delta{\cal R}_u(\LLRc)}}
\newcommand{\Lqcd}{\ensuremath{\Lambda_{\mathrm{QCD}}}}
\newcommand {\breco}{\ensuremath{B_{\mathrm{r}}}}
\newcommand {\brecoil}{\ensuremath{\Bb_{\mathrm{sl}}}}
\newcommand {\lone}{\ensuremath{\lambda_1}}
\newcommand {\pmiss}{\ensuremath{p_{\mathrm{miss}}}}
\newcommand {\vpmiss}{\ensuremath{{ \textbf{p} }_{\mathrm{miss}}}}
\newcommand {\mmiss}{\ensuremath{m_{\mathrm{miss}}^2}}
\newcommand {\Nsl}{\ensuremath{N_{\mathrm{sl}}}}
\newcommand {\mES}{\ensuremath{m_{\mathrm{ES}}}}
\def\ie                 {{\it i.e.}}

\newcommand{\vubts} {\ensuremath {\Vub/\Vts}}

\RequirePackage{xspace}

\long\def\inst#1{\par\nobreak\kern 4pt\nobreak
    {\it #1}\par\vskip 10pt plus 3pt minus 3pt}

\def\varia1c-a{666}

\begin{document}

 \preprint{\babar-PUB-\BABARPubYear/\BABARPubNumber}
 \preprint{SLAC-PUB-\SLACPubNumber}
 
 \begin{flushleft}
 \babar-PUB-\BABARPubYear/\BABARPubNumber\\
 SLAC-PUB-\SLACPubNumber\\
 hep-ex/\LANLNumber\\
 \end{flushleft}
 
 \begin{flushright}
 \end{flushright}

\title[Short Title] {Determinations of \Vub\ from Inclusive Semileptonic $B$
  Decays with Reduced Model Dependence}

\author{B.~Aubert}
\author{R.~Barate}
\author{D.~Boutigny}
\author{F.~Couderc}
\author{Y.~Karyotakis}
\author{J.~P.~Lees}
\author{V.~Poireau}
\author{V.~Tisserand}
\author{A.~Zghiche}
\affiliation{Laboratoire de Physique des Particules, F-74941 Annecy-le-Vieux, France }
\author{E.~Grauges}
\affiliation{IFAE, Universitat Autonoma de Barcelona, E-08193 Bellaterra, Barcelona, Spain }
% ...
\collaboration{The \babar\ Collaboration}
\noaffiliation

\begin{abstract}
We report two novel determinations of \Vub\ with reduced model
dependence, based on measurements of the mass distribution of the
hadronic system in semileptonic $B$ decays.  Events are selected by
fully reconstructing the decay of one $B$ meson and identifying a
charged lepton from the decay of the other $B$ meson from \upsbb\
events.  In one approach, we combine the inclusive \Bxulnu\ rate,
integrated up to a maximum hadronic mass $\mX < 1.67\gevcc$, with a
measurement of the inclusive \Bsg\ photon energy spectrum.  We obtain
$\Vub = (4.43 \pm 0.38_{\mathrm{stat}} \pm 0.25_{\mathrm{syst}} \pm
0.29_{\mathrm{theo}} ) \times 10^{-3}$.  In another approach we
measure the total \Bxulnu\ rate over the full phase space and find
$\Vub = (3.84 \pm 0.70_{\mathrm{stat}} \pm 0.30_{\mathrm{syst}} \pm
0.10_{\mathrm{theo}} ) \times 10^{-3}$.

\end{abstract}


\date{\today} 

\maketitle

The measurement of the element $V_{ub}$ of the Cabibbo-Kobayashi-Maskawa
quark-mixing matrix~\cite{ckm} plays a critical role in testing the
consistency of the Standard Model description of \CP\ violation. The
uncertainties in existing measurements~\cite{ref:oldmx,ref:previous} are
dominantly due to uncertainties in the $b$-quark mass $m_b$ and the modeling
of the Fermi motion of the $b$ quark inside the \Bb\
meson~\cite{fermimotion}.  In this paper, we present two techniques to
extract \Vub\ from inclusive \Bxulnu~\cite{cc} decays where these
uncertainties are significantly reduced. Neither method has been previously
implemented experimentally.

Leibovich, Low, and Rothstein (LLR) have presented a prescription to extract
\Vub\ with reduced model dependence from either the lepton energy or the
hadronic mass \mX~\cite{LLR}.  A technique utilizing weight functions had
been proposed previously by Neubert~\cite{fermimotion}. The calculations
of LLR are accurate up to corrections of order $\alpha_s^2$ and
$(\Lambda m_B /(\LLRc\,m_b))^2$, where $\LLRc$ is the experimental maximum
hadronic mass up to which the \Bxulnu\ decay rate is determined and $\Lambda
\approx \Lqcd$. This method combines the hadronic mass spectrum, integrated
below \mXmax , with the high-energy end of the measured differential \Bsg\
photon energy spectrum via the calculations of LLR.

An alternative method~\cite{Uraltsev:1999rr} to reduce the model dependence
is to measure the \Bxulnu\ rate over the entire \mX\ spectrum.  Since no
extrapolation is necessary to obtain the full rate, systematic uncertainties
from $m_b$ and Fermi motion are much reduced.  Perturbative corrections are
known to order $\alpha_s^2$. We extract the \Bxulnu\ rate from the hadronic
mass spectrum up to $\mXmax = 2.5\gevcc$ which corresponds to about 96\% of
the simulated hadronic mass spectrum.

The measurements presented here are based on a sample of 88.9 million \BB
pairs collected near the \FourS\ resonance by the \babar\
detector~\cite{babarnim} at the PEP-II asymmetric-energy $e^+e^-$ storage
rings operating at SLAC.  The analysis uses \FourS\to\BB\ events in which one
of the \B mesons decays hadronically and is fully reconstructed (\breco) and
the other decays semileptonically (\brecoil).  To reconstruct a large sample
of $B$ mesons, we follow the procedure described in Ref.~\cite{ref:oldmx} in
which charged and neutral hadrons are combined with an exclusively
reconstructed $D$ meson to obtain combinations with an energy consistent with
a $B$ meson.  While this approach results in a low overall event selection
efficiency, it allows for the precise determination of the momentum, charge,
and flavor of the \breco\ candidates.

%
%
We use Monte Carlo (MC) simulations of the \babar\ detector based on
\geantfour~\cite{geant} to optimize selection criteria and to determine
signal efficiencies and background distributions.  Charmless semileptonic
\Bxulnu\ decays are simulated as a combination of resonant three-body decays
($X_u = \pi, \rho, \omega, \eta, \eta^{\prime}$)~\cite{ref:isgwtwo}, and
decays to non-resonant hadronic final states $X_u$~\cite{ref:fazioneubert}
for which the hadronization is performed by \jetset~\cite{ref:jetset}. The
effect of Fermi motion is implemented in the simulation using an exponential
function~\cite{ref:fazioneubert} with the parameters $m_b = 4.79\gevcc$ and
$\lone=-0.24\gev^2\!/c^4$~\cite{ref:babarellipse}. The simulation of the
\Bxclnu\ background uses a Heavy Quark Effective Theory parameterization of
form factors for $\Bb\to D^{*}\ell\nub$~\cite{ref:FF} and models for $\Bb\to
D \pi \ell\nub, D^* \pi \ell\nub$~\cite{ref:goityroberts} and $\Bb\to D
\ell\nub,D^{**}\ell\nub $~\cite{ref:isgwtwo} decays.

Semileptonic \brecoil\ candidates are identified by the presence of at least
one electron or muon with momentum $p_\ell^* > 1\gevc$ in the \brecoil\ rest
frame.  For charged \breco\ candidates, we require the charge of the lepton
to be consistent with a primary decay of a \brecoil . For neutral \breco\
candidates, both charge-flavor combinations are retained and the average
$\Bz$-$\Bzb$ mixing rate~\cite{ref:pdg2004} is used to determine the primary
lepton yield.  Electrons (muons) are identified~\cite{Aubert:2002uf}
(Ref.~\cite{babarnim}), with a 92\% (60--75\%) average efficiency and a
hadron misidentification rate ranging between 0.05\% and 0.1\% (1--3\%).

The hadronic system $X$ in the \Bxlnu\ decays is reconstructed from charged
tracks and energy depositions in the calorimeter that are not associated with
the \breco\ candidate or the identified lepton.  The neutrino four-momentum
$p_{\nu}$ is estimated from the missing momentum four-vector $\pmiss =
p_{\FourS}-p_{\breco} -p_X-p_\ell$, where all momenta are measured in the
laboratory frame and $p_{\FourS}$ refers to the \FourS\ momentum.

To select \Bxulnu\ candidates we require exactly one lepton with $p_\ell^* >
1\gevc$ in the event, charge conservation ($Q_{X} + Q_\ell + Q_{\breco} =
0$), and a missing four-momentum consistent with a neutrino hypothesis, \ie ,
missing mass consistent with zero ($-1.0 < \mmiss < 0.5\,\gev^2/c^4$),
$|\vpmiss| > 0.3\gevc$, and $|\cos\theta_\mathrm{miss}| < 0.95$, where
$\theta_\mathrm{miss}$ is the polar angle of the missing momentum
three-vector \vpmiss.  These criteria suppress the majority of \Bxclnu\
decays that contain additional neutrinos or an undetected $\KL$
meson. Additionally we reject events with charged or neutral kaons
(reconstructed as $\KS\to\pi^+\pi^-$ decays) in the decay products of the
\brecoil .  We suppress $\Bb\to\Dstar\ell\overline{\nu}$ backgrounds by
partial reconstruction of charged and neutral $D^*$ mesons via identification
of charged and neutral slow pions.  The reconstruction of the mass of the
hadronic system is improved by a kinematic fit that imposes four-momentum
conservation, the equality of the masses of the two $B$ mesons, and
$p_{\nu}^2 = 0$.  The resulting \mX\ resolution is $\sim250\mevcc$ on
average.

\begin{figure}[t]
  \begin{centering} 
    \hbox{\hskip-0.cm \epsfig{file=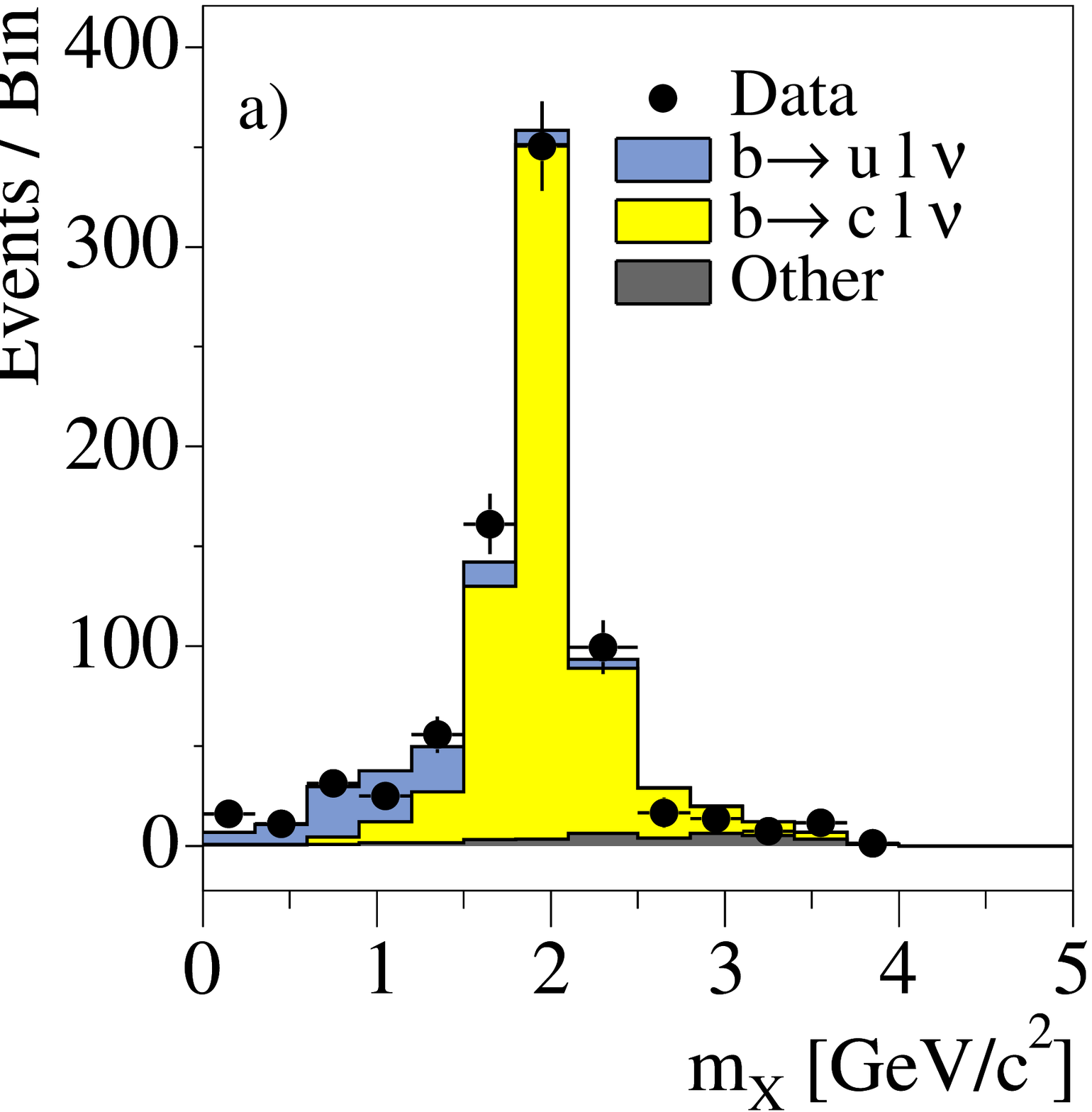,height=4.6cm} \hskip-0.3cm \epsfig{file=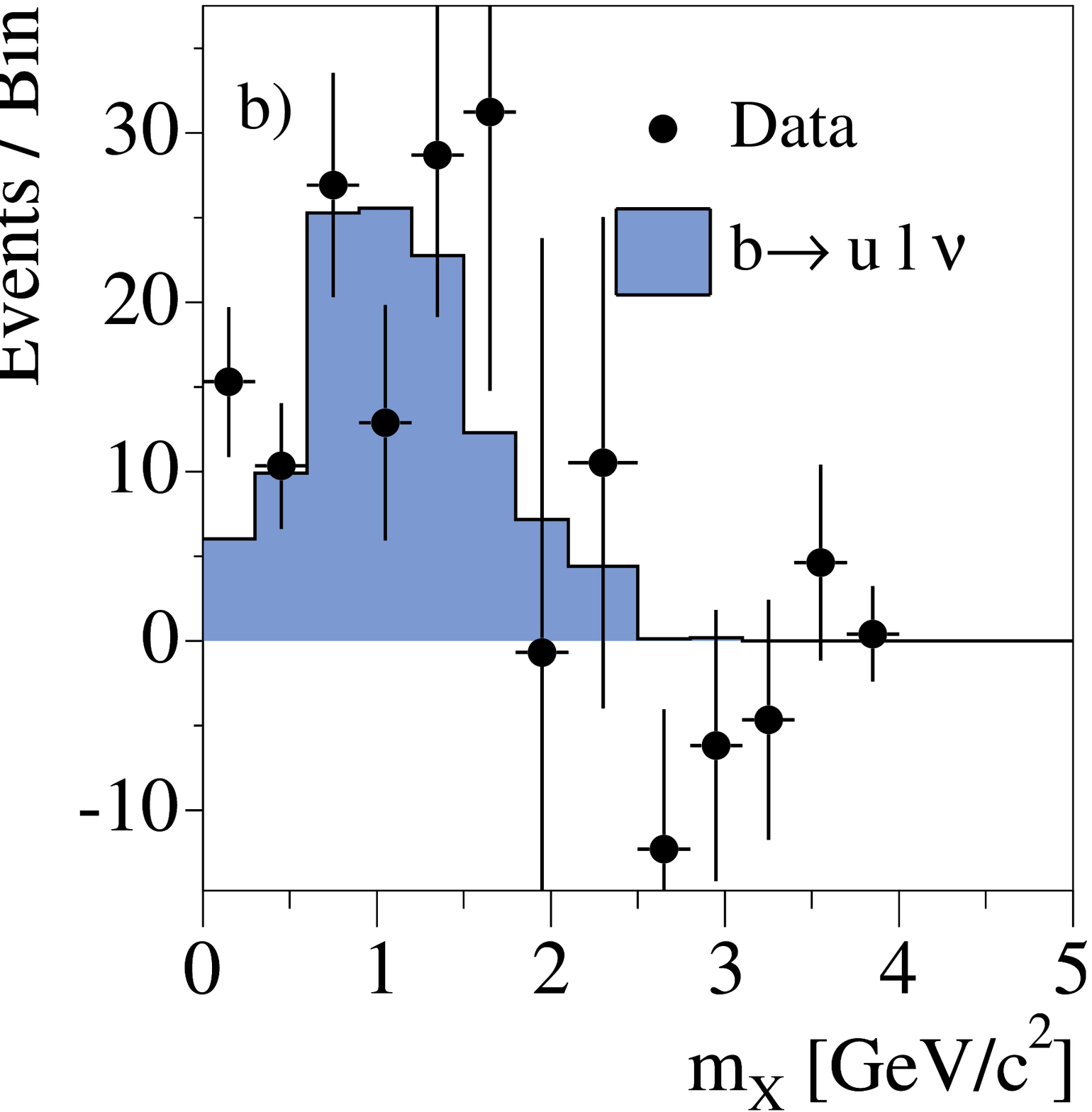,height=4.6cm}}
    \caption{ The \mX\ distributions (without combinatorial backgrounds) for
      \Bxlnu\ candidates: a)~data (points) and fit components after the
      minimum-$\chi^2$ fit, and b)~data and signal MC after subtraction of
      the \Bxclnu\ and other backgrounds. The upper edge of the eighth bin is
      chosen to be at $\mX = 2.5\gevcc$. This fit result, with $\chi^2 =
      10.2$ for 11 degrees of freedom, is used to extract the number of
      signal events below $2.5\gevcc$.
      \label{fig:mxLLR}}
  \end{centering}
\end{figure}

%
%
The extraction of \vubts\ from the selected events starts from the
equation~\cite{LLR}
\begin{equation}
  \frac{\Vub}{\Vts} =
  \bigg\{\frac{6\alpha(1+H_{\mathrm{mix}}^\gamma)(C_7^{(0)})^2}{\pi[I_0(\LLRc)+
  I_+(\LLRc)]} \times \dGc \bigg\}^{\frac{1}{2}} ,
  \label{e:expZero}
\end{equation}
\noindent where \dGc\ is the partial charmless semileptonic decay rate
extracted from the number of \Bxulnu\ events up to a limit \LLRc\ in the \mX\
spectrum. $H_{\mathrm{mix}}^\gamma$ accounts for interferences between
electromagnetic penguin operator $O_7$ with $O_2$ and
$O_8$~\cite{Greub:1996jd}, and $C_7^{(0)}$ is the effective Wilson
coefficient.  The terms $I_0(\LLRc)$ and $I_+(\LLRc)$ are determined by
multiplying the photon energy spectrum $d\Gamma^\gamma / d E_\gamma$ in \Bsg\
decays~\cite{ref:babarellipse} with weight functions~\cite{LLR} and
integrating. The weights are zero below a minimum photon energy
$E_\gamma^{\mathrm{min}} = m_B/2 - \mXmax/4$.

In terms of measurable quantities, \dGc\ is
\begin{equation}
  \dGc = \frac{ N_u(\LLRc) f(\LLRc) {\cal B}(\Bxlnu)}{ \Nsl \varepsilon_u(\LLRc) } 
  \times
  \frac{\varepsilon^{\mathrm{sl}}_{\ell}}{\varepsilon^{u}_{\ell}}
  \times
  \frac{\varepsilon^{\mathrm{sl}}_{\mathrm{reco}}}{\varepsilon^{u}_{\mathrm{reco}}} .
  \label{e:expOne}
\end{equation}
\noindent Here, $N_u(\LLRc)$ is the number of reconstructed \Bxulnu\ events
with $\mX < \mXmax$, $f(\LLRc)$ accounts for migration in and out of the
region below \mXmax\ due to finite \mX\ resolution, ${\cal B}(\Bxlnu)$ is the
total inclusive semileptonic branching fraction, and $\varepsilon_{u}(\LLRc)$
is the efficiency for selecting \Bxulnu\ decays once a \Bxlnu\ decay has been
identified with a hadronic mass below \mXmax . \Nsl\ is the number of
observed fully reconstructed \B meson decays with a charged lepton with
momentum above $1\gevc$, $\varepsilon^{\mathrm{sl}}_{\ell} /
\varepsilon^{u}_{\ell}$ corrects for the difference in the efficiency of the
lepton momentum selection for \Bxlnu\ and \Bxulnu\ decays, and
$\varepsilon^{\mathrm{sl}}_{\mathrm{reco}} / \varepsilon^{u}_{\mathrm{reco}}$
accounts for the difference in the efficiency of reconstructing a \breco\ in
events with a \Bxlnu\ and \Bxulnu\ decay.  By measuring the ratio of \Bxulnu\
events to all semileptonic $B$ decays many systematic uncertainties cancel
out.

We derive $N_u(\LLRc)$ from the \mX\ distribution with a binned $\chi^2$ fit
to four components: data, \Bxulnu\ signal MC, \Bxclnu\ background MC, and a
small MC background from other sources (misidentified leptons, \Btaunu , and
charm decays), fixed relative to the \Bxclnu\ component.  $N_u(\LLRc)$ is
determined after the subtraction of the fitted background contributions.  For
all four contributions, the combinatorial background is determined,
separately in each bin of the \mX\ distribution, with unbinned maximum
likelihood fits to distributions of the beam energy-substituted mass $\mES =
\sqrt{s/4-{\bf p}_B^{\;2}}$ of the \breco\ candidate, where $\sqrt{s}$ is the
$e^+e^-$ center-of-mass energy. The \mES\ fit uses an empirical description
of the combinatorial background shape~\cite{argusf} with a signal
shape~\cite{cry} peaking at the $B$ meson mass.  The combinatorial background
varies from 5\% (low \mX\ bins) to 25\% (high \mX\ bins).  The fitted $\mX$
distributions are shown in Fig.~\ref{fig:mxLLR} before (a) and after (b)
subtraction of backgrounds.  The \mX\ bins are $300\mevcc$ wide except that
one bin is widened such that its upper edge is at \mXmax.

We extract $\Nsl = (3.253\pm0.024) \times 10^4$ from an unbinned maximum
likelihood fit to the \mES\ distribution of all events with $p_\ell^* >
1\gevc$.  The efficiency corrections $\varepsilon^{\mathrm{sl}}_{\ell} /
\varepsilon^{u}_{\ell} = 0.82 \pm 0.02_\mathrm{stat}$, as well as
$\varepsilon_u(\LLRc)$ and $f(\LLRc)$ (see Table~\ref{tab:measurements}) are
derived from simulations, where we also find
$\varepsilon^{\mathrm{sl}}_{\mathrm{reco}} / \varepsilon^{u}_{\mathrm{reco}}$ in
agreement with one, assigning a 3\% uncertainty.

\begin{table}
  \begin{center}
    \caption{ Quantities in Eq.~\ref{e:expOne} that depend on \mXmax\
      and their statistical uncertainties.  The LLR (full rate)
      technique is given in the first (second) column.}
    \vspace{0.03in}
    \begin{tabular}{p{25mm}rclp{5mm}rcl}
      \hline
      \hline     
      \mXmax             & \multicolumn{3}{c}{$1.67\gevcc$ } & &
      \multicolumn{3}{c}{$2.50\gevcc$ }  \\
      \hline     
      $f$                & $1.010$&$\pm$&$0.005$ && $0.998$&$\pm$&$0.002$ \\
      $N_u$              & $120$&$\pm$&$17$      && $135$&$\pm$&$45$      \\
      $\varepsilon_u$       & $0.231$&$\pm$&$0.005$ && $0.231$&$\pm$&$0.004$ \\
      $\delta{\cal R}_u \times 10^3$ & $1.43$&$\pm$&$0.21$   && $1.59$&$\pm$&$0.53$   \\
      \hline
      \hline
    \end{tabular}
    \label{tab:measurements}
  \end{center}
\end{table}

\begin{figure}[t]
  \begin{centering} 
    \epsfig{file=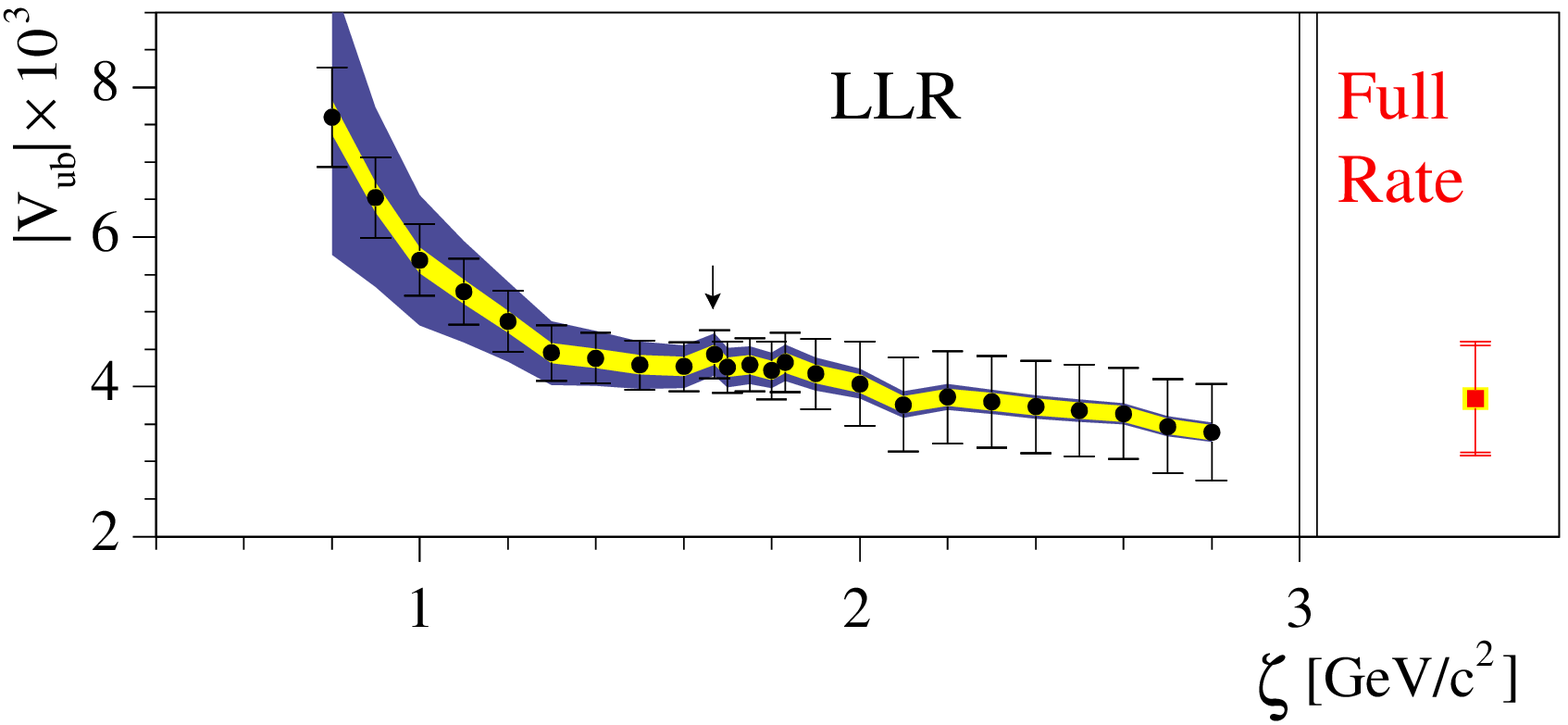,height=4cm}
    \caption{\Vub\ as a function of \mXmax\ with the LLR method (left) and
      for the determination with the full rate measurement (right).  The
      error bars indicate the statistical uncertainty. They are correlated
      between the points and get larger for larger \mXmax\ due to larger
      background from \Bxclnu .  The total shaded area illustrates the
      theoretical uncertainty; the inner light shaded (yellow) area indicates
      the perturbative share of the uncertainty. The arrow indicates
      $\mXmax=1.67\gevcc$.
      \label{fig:mxscan}
    }
  \end{centering}
\end{figure}

%
%

%
%
We study three categories of systematic uncertainties in the determination of
\Vub : uncertainties in the signal extraction, the simulation of physics
processes, and the theoretical description. The quoted uncertainties have
been determined for a value of $\mXmax = 1.67\gevcc$ where the total
uncertainty on \Vub\ is found to be minimal.

Experimental uncertainties in the signal extraction arise from imperfect
description of data by the detector simulation.  We assign 0.5\% (0.5\%,
0.8\%) for the particle identification of electrons ($\mu$, $K^\pm$), 0.7\%
for the reconstruction efficiency of charged particles, and 0.8\% for the
resolution and reconstruction efficiency of neutral particles. An additional
0.9\% uncertainty is due to imperfect simulation of \KL\ interactions. By
changing the function describing the signal shape in \mES\ to a Gaussian
function and switching from an unbinned to a binned fit method we derive an
uncertainty of 2.2\%.  An uncertainty of 0.8\% is determined by letting the
contribution from other sources (see above) to the \mX\ spectrum float freely in the
minimum-$\chi^2$ fit. The uncertainties on the inclusive \Bsg\ photon energy
spectrum are propagated including the full correlation matrix between the
individual bins.

%
%
The second category of systematic uncertainties arises from imperfections in
the composition and dynamics of decays in the simulation, both in signal and
background.  The uncertainties in the branching fractions of $B\to D^{(*,
**)}l\bar{\nu} X$ decays~\cite{ref:pdg2004} contribute 0.7\%. The
uncertainties in the form factors in $B\to D^{*}l\bar{\nu}$
decays~\cite{ref:FF} introduce a 0.3\% uncertainty.  Branching fractions of
$D$-meson decay channels~\cite{ref:pdg2004} contribute 0.2\%.  The relative
contribution of the non-resonant final states has been varied by 20\%
resulting in an uncertainty of 0.5\%.  The branching fractions of the
resonant final states have been varied by $\pm 30\%$ ($\pi$, $\rho$), $\pm
40\%$ ($\omega$), and $\pm 100\%$ ($\eta$ and $\eta^{\prime}$ simultaneously)
resulting in an uncertainties of 1.0\%.  An uncertainty of 0.7\% due to
imperfect description of hadronization is determined from the change observed
when we saturate the spectrum with the non-resonant component alone.  We
derive a 1.3\% uncertainty due to the imperfect modeling of the
$K\overline{K}$ content in the $X_u$ system by varying the fraction of decays
to $s\bar{s}$-pairs by 30\% for the non-resonant
contribution~\cite{Althoff:1984iz}.  Even though the extraction of \Vub does
not explicitly depend on a model for Fermi motion, there is still a residual
dependency via the simulation of signal events.  By varying the Fermi motion
parameters $m_b$ and \lone\ within their respective uncertainties, taking
correlations into account~\cite{ref:babarellipse}, we derive an uncertainty
of 3.5\%.

%
%

We calculate theoretical uncertainties in the weighting technique by varying
the input parameters and repeating the weighting procedure including the
calculation of all variables: $H_{\mathrm{mix}}^{\gamma}$, $\alpha_S$, and
Wilson-coefficients.  We vary $\alpha$ between $\alpha (m_b)$ and
$\alpha(m_W)$ with a central value of $1/130.3$ and find an uncertainty of
less than $1\%$. For perturbative effects, an uncertainty of 2.9\% is derived
by varying the renormalization scale $\mu$ between $ m_b/2$ and $2m_b$.
Non-perturbative effects are expected to be of the order $(\Lambda m_B
/(\LLRc\,m_b))^2$, where $\Lambda = 500\mevcc$~\cite{LLRpriv}, resulting in
an uncertainty of 5.4\%.
Theoretical uncertainties in the measurement via the full rate are taken from
Ref.~\cite{Buchmuller:2005zv} to be $1.2\%$ (QCD) and $2.2\%$ (HQE).
Table~\ref{tab:breakdown} provides a summary of the uncertainties for $\mXmax
= 1.67\gevcc$ and for $\mXmax = 2.5\gevcc$.

\begin{table}
  \begin{center}
    \caption{Summary of results and uncertainties on \Vub for both approaches. The LLR
      (full rate) technique is given in the first (second) column. }
    \vspace{0.03in}
    \begin{tabular}{lp{8mm}cp{5mm}c} 
      \hline
      \hline
      $\mXmax\ [\gevcc \, ]$            & &      1.67           & &     2.5 \\
      \hline                                                      
      $\Vub \times 10^{3}$             & &      4.43           & &     3.84           \\
      \hline                                                      
      \Bxulnu\ stat.                   & &      7.7\%          & &    18.2\%          \\
      experimental syst.               & &      3.3\%          & &     3.6\%          \\
      background model                 & &      1.0\%          & &     3.8\%          \\
      signal model                     & &      3.9\%          & &     5.6\%          \\
      theoretical                      & &      6.2\%          & &     2.6\%          \\
      \Bsg\ (stat., syst.)             & &      3.5\%, 2.0\%   & &     ---            \\
      \Vcb\ (exp., theo.)              & &      1.0\%, 1.7\%   & &     ---            \\
      \hline
      \hline
    \end{tabular}
    \label{tab:breakdown}
  \end{center}
\end{table}

%
%
Finally, we present two different determinations of \Vub . First, using the
weighting technique with the photon energy spectrum in \Bsg\ decays from
Ref.~\cite{ref:babarellipse}, the hadronic mass spectrum up to a value of
$\mXmax = 1.67\gevcc$, we find $|V_{ub}|/|V_{ts}| = 0.107 \pm
0.009_{\mathrm{stat}} \pm 0.006_{\mathrm{syst}} \pm 0.007_{\mathrm{theo}}$.
If we assume the CKM matrix is unitary then $\Vts = \Vcb \times (1\pm
{\cal O}(1\%))$ and, taking \Vcb from Ref.~\cite{Aubert:2004aw}, we derive
\begin{equation*}
  \Vub = (4.43 \pm 0.38 \pm 0.25 \pm 0.29 ) \times 10^{-3} ,
\end{equation*}
where the first error is the statistical uncertainty from \Bxulnu\ and from
\Bsg\ added in quadrature, the second (third) is systematic (theoretical).
Second, we determine \Vub from a measurement of the full \mX\ spectrum, \ie,
up to a value of $\mXmax = 2.5\gevcc$, and find $\Vub = (3.84 \pm
0.70_{\mathrm{stat}} \pm 0.30_{\mathrm{syst}} \pm 0.10_{\mathrm{theo}} )
\times 10^{-3}$, using the average $B$ lifetime of $\tau_B = (1.604 \pm
0.012)\ps$~\cite{ref:pdg2004,ref:b0bch}.


The weighting technique is expected to break down at low values of \mXmax ,
since only a small fraction of the phase space is used.
Figure~\ref{fig:mxscan} illustrates the dependence of the result, and its
statistical and theoretical uncertainties, on variations of \mXmax\ and also
compares it with the value of \Vub\ determined from the full rate. The
weighting technique appears to be stable down to $\mXmax \sim 1.4\gevcc$.
The current uncertainties on the \Bsg\ photon energy spectrum limit the
sensitivity with which the behavior at high \mXmax\ can be probed.

The above results are consistent with previous
measurements~\cite{ref:oldmx,ref:previous} but have substantially smaller
uncertainties from $m_b$ and the modeling of Fermi motion.  Both techniques
are based on theoretical calculations that are distinct from other
calculations normally employed to extract \Vub and, thus, provide a
complementary determination of \Vub.


\begin{acknowledgments}
We wish to thank Adam Leibovich, Ian Low, and Ira Rothstein for their help
and support. We are grateful for the excellent luminosity and machine
conditions provided by our \pep2\ colleagues, and for the substantial
dedicated effort from the computing organizations that support \babar.  The
collaborating institutions wish to thank SLAC for its support and kind
hospitality.  This work is supported by DOE and NSF (USA), NSERC (Canada),
IHEP (China), CEA and CNRS-IN2P3 (France), BMBF and DFG (Germany), INFN
(Italy), FOM (The Netherlands), NFR (Norway), MIST (Russia), and PPARC
(United Kingdom).  Individuals have received support from the A.~P.~Sloan
Foundation, Research Corporation, and Alexander von Humboldt Foundation.
\end{acknowledgments}

%
%
%
%
%
%

\label{sec:references}


\begin{thebibliography}{02}
\bibitem{cc}
Charge conjugation is implied throughout the letter.
\bibitem{ref:b0bch} The impact of the uncertainty of the relative fraction of
produced neutral and charged $B$ mesons is negligible.
\end{thebibliography}
\end{document}